\documentstyle[12pt]{article}
\newcommand{\ber}{\begin{eqnarray}}
\newcommand{\eer}{\end{eqnarray}}
\newcommand{\bea}{\begin{equation}}
\newcommand{\eea}{\end{equation}}
\begin{document}
\title{CRITICAL ULTRASONICS NEAR THE SUPERFLUID TRANSITION : FINITE SIZE 
EFFECTS}
\author{Saugata Bhattacharyya and J. K. Bhattacharjee \\
Department of Theoretical Physics \\
Indian Association for the Cultivation of Science \\
Jadavpur, Calcutta 700 032, India}
\date{}
\maketitle
\begin{abstract}
The suppression of order parameter fluctuations at the boundaries causes the 
ultrasonic attenuation near the superfluid transition to be lowered below the bulk value . We calculate explicitly the first deviation from the bulk value for temperatures above the lambda point . This deviation is significantly larger than for static quantities like the thermodynamic specific heat or other transport properties like the thermal conductivity .
This makes ultrasonics a very effective probe for finite size effects . \\\\\\
PACS number(s): 64.60.Ht
\end{abstract}
\newpage
Critical phenomena in confined geometry has been attracting a fair amount of 
attention of late because of the progress on the experimental front$[2-9]$
which is making it possible to check the predictions of finite size effects 
(FSE). A fair amount of this experimental effort has gone in studying the 
specific heat near the superfluid transition . With the bulk specific heat 
quite well understood and the existence of a sharp phase transition (apart from gravity rounding , which too can be removed by doing experiments in space) established , efforts have been made to study the FSE . It is expected that the FSE will round out the transition and hence the divergence at $T=T_{\lambda}$ will be removed . The specific heat will be finite and the finite value will be a function of the confining length . We will keep in mind  one of the favoured experimental geometries, where one takes two parallel plates separated by a distance $L$ , much smaller than the linear dimensions of the plates . For $L\gg\xi$ , the correlation length at a given temperature, usual thermodynamic result follows . It is when $L\leq\xi$ , that FSE dominate . Finite size scaling suggests the existence of a scaling function , function of $\xi/L$- in terms of which the theory can be cast . The specific heat $C(t,L)$ in finite geometry has the form $C(t,L)\sim t^{-\alpha} g(t^{-\nu}/L)+ Constant$ , where $\xi\sim t^{-\nu}$ and $t=(T-T_{\lambda})/T_{\lambda}$ , $T_{\lambda}$ being the transition temperature . The function {\bf{g}} has been calculated by various authors $[10-12]$ . In what follows we propose a method of checking for FSE by studying a related dynamic property . This is the study of ultrasonic attenuation (UA) near $T_{\lambda}$ at high frequencies . In fact, it is our contention that UA is one of the best ways of checking for FSE since the single surface effect alone can produce effects greater than $10\%$ . The critical 
fluctuations relax according to $\xi^{-z}$, where z is the dynamic scaling 
exponent . For frequencies $\omega$ such that $\omega\gg \Gamma_{0}$ (occurs if one is close to the critical point) , the attenuation is independent of the correlation length. For a finite size system , we will show below that in this limit , the attenuation is determined by $\omega$ and $L$ alone .  
We provide explicit answers for frequencies $\omega$ which are much smaller
than a cutoff frequency $\omega_0$ (of the order of a few GHz) and for plate 
separation $L\geq(2\Gamma_0/\omega)^{\frac{1}{2}}$ for a given frequency 
$\omega$ . Our prediction for the attenuation per wavelength as a function of $ \omega$ for the plate separation of $2110\AA$ is shown in Fig(1) . It should be possible to check the prediction experimentally . In fact , this should be the simplest way of checking for FSE since the effect is quite pronounced (about $18\%$ at $L=2110\AA$ and $\omega=10 Mhz$) for the available  confining lengths as shown in Fig(1) . This occurs because the imaginary part of the specific heat determines the UA and which is much smaller than the real part , but as we shall see below , {\bf{both are equally affected by the FSE}} . Consequently, the relative effect is much larger for the imaginary part and this will show up in the UA .
\par
The basis of our calculation is once more the Pippard Buckingham Fairbank (PBF) relation $[13-14]$ which gives a successful account $[15-17]$ of the critical ultrasonics in the situation where $L\gg \xi$ . The PBF relation is obtained from general considerations of entropy clamping and yields for the sound velocity $u(T,\omega)$
\bea            
u(T,\omega)= u_{0}(T_{0})+u_{1}C_{0}/C_{P}(T,\omega)                   
\eea
where $u_{0}(T_{0})$ is the sound speed at the transition point ($T_{0}$ is the bulk $T_{\lambda}$ for the infinite system , but is a L-dependent temperature for the finite size system) , $u_{1}$ and $C_{0}$ are constants and $C_{P}(T,\omega)$ is the specific heat at finite frequency . 
\par
For the bulk case , $C_{P}(T,\omega=0)$ diverges at $T=T_{\lambda}$ and 
$C_{P}(T,\omega)$ is a homogeneous function of $\omega$ and $\xi$ . If the characteristic relaxation rate is $\Gamma_{0}\xi^{-z}$ then the scaling form of $C_{P}^{bulk}$ is
\bea
C_{P}^{bulk}(T,\omega)=\xi^{\alpha/\nu} f(\frac{\omega}{\Gamma_0\xi^{-z}})     
\eea
The exponent $\alpha$ is very close to zero for the superfluid transition in in ${}^4He$ and for many practical purposes , it is possible to write
\bea
C_{P}^{bulk}(T,\omega)=C[\ln(\Lambda\xi)+f(\frac{\omega}{\Gamma_0\xi^{-z}})]   
\eea
The function $f(\omega/(\Gamma_0\xi^{-z}))$ reduces to a constant for $\omega=0$ and tends to $-\ln(\omega/\Gamma_0)^{1/z}\xi$ for $\omega\gg\Gamma_0\xi^{-z}$ . A one loop calculation of the scaling function $f(\Omega)$ where $\Omega=\frac{\omega}{\Gamma_0\xi^{-z}}$ , was carried out and led to a successful scaling theory of the attenuation in the bulk ${}^4He$ near $T_{\lambda}$ $[15-17]$ .
\par
We now need to discuss the effect of a confining geometry . At zero frequency , the specific heat is blunted due to the FSE and the usually divergent specific heat remains finite . The single loop calculation of the scaling function $g(\xi/L)$ discussed before gives a very reasonable account of the recent specific heat data by Mehta and Gasparini $[2]$ . One of the most important feature of the scaling function is the low $\xi/L$ limit (experimentally most easily accessible) is the first departure from the thermodynamic limit - the magnitude of this departure $\Delta C$ has to be proportional to the surface (A) to volume (V) ratio and hence from purely dimensional arguments , the correction can be written as
\bea
\Delta C = C(\xi,L)- C_{\infty}(\xi)= - a C A\frac{\xi}{V}
\eea
where $a$ is a number of $O(1)$ , that can be obtained from the function
$g(\xi/L)$ , and $C$ is the dimensional constant defined in eqn(3) .  The value of 
$a$ as inferred from Scmolke et al. [11] is 1.4.
The agreement of this departure with the measured departure of Mehta and 
Gasparini is impressive .
\par
In our present concern we need the three variable function $C(\xi,L,\omega)$ , whose two limits $C(\xi,\omega)$ and $C(\xi,L)$ are already well known . We will characterize $C(\xi,L,\omega)$ by its first departure from the infinite volume limit $C(\xi,\omega)$ and write the generalization of eqn(4) as
\bea
\Delta C(\xi,L,\omega)=C(\xi,L,\omega)-C(\xi,\omega)=- a(\xi,\omega)C(\xi) A/V
\eea
where $a(\xi,\omega)$ is a scaling function , whose zero frequency limit is 
$a\xi$ (see Eqn(4)) and whose general form will be presented below . As soon 
as we start discussing the scaling function for $C(\xi,L,\omega)$ we need to 
worry about what sets the scale for $\omega$ . As we have discussed above , this has to be the rate of decay of fluctuations $\Gamma(\xi)$. In the finite geometry that we are discussing now , the scale for decay of fluctuations will depend on $L$ as well . In discussing the correction depicted in Eqn(5) , it is obvious that this fine point need not be discussed as this correction is already $O(\frac{1}{L})$ . For He (superfluid transition) , there is in someways an additional simplifying feature . For the order parameter decay rate the non-linear effect of fluctuations becomes significant , only very close to the critical point and for all practical purposes , the relaxation rate can be taken to be at its non critical background value .
\par
The complex order parameter field $\psi_{i}(x) {i=1,2}$ will be governed by the Langevin equation
\bea
\dot{\psi_i}=-\Gamma_0\frac{\delta F}{\delta {\psi_i}}+ N_i
\eea
where
\bea
F=\int d^{D}x  [ \frac{m^2}{2}\psi^{2}+\frac{1}{2}(\nabla\psi)^{2}
+\frac{\lambda}{4}(\psi^{2})^{2} ]
\eea
and N is a Gaussian white noise . For reasons stated above we choose to drop the reversible term (the Josephson equation for the phase of the order parameter) . The parameter $m^{2}$ is proportional to $T-T_{\lambda}$, where $T_{\lambda}$ is the bulk transition temperature . The system is confined in one of the D directions . We call that the z-direction . It is convenient to work with the fourier transform in $D-1$ directions and the fourier series (Dirichlet boundary conditions at $z=0$ and $z=L$ suppressing the fluctuations) in the z direction . The expansion of the time-dependent order parameter field is
\bea
\psi_i(\vec{r},t)=\sum_{n}\psi_i(n,K,t)\exp^{i\vec{K}.\vec{R}}\sin(\frac
{n\pi\!z}{L})
\eea
The equation of motion for $\psi_i(n,K,t)$ is
\bea
\dot{\psi_i}(n,K,t)=-\Gamma_0(m^{2}+K^{2}+\frac{n^{2}\pi^{2}}{L^{2}})\psi_i
(n,K,t)+N_i+O(\psi^{3})
\eea
In what follows , we will assume that all static correlations have been
accounted for and $m^{2}=\xi^{-2}$ . The specific heat is obtained as the response function corresponding to the time dependent correlation function
\bea
D(\xi,L,t_{12})=\int\!\int\!\int dz_1 dz_2 d^{D}R_{12}<\psi^{2}(\vec
{R_1},z_1,t_1)\psi^{2}(\vec{R_2},z_2,t_2)>
\eea
with $D(\xi,L,\omega)=2 \frac{Im C(\xi,L,\omega)}{\omega}$ according to
fluctuation dissipation theorem , straightforward algebra leads to (two term 
accuracy , $L\rightarrow\infty$ limit and the first correction)
\ber
C(\xi,L,\omega)&=&\int\frac{d^{D}p}{(2\pi)^{D}}\frac{1}{(p^{2}+m^{2})}
\frac{1}{(-\frac{i\omega}{2\Gamma_0}+p^{2}+m^{2})} \nonumber \\
& &-\frac{1}{2L}\int\frac{d^{D-1}p}{(2\pi)^{D-1}}\frac{1}{(p^{2}+m^{2})}
\frac{1}{(-\frac{i\omega}{2\Gamma_0}+p^{2}+m^{2})}
\eer
We work to logarithmic accuracy and hence evaluate the integrals at $D=4$
(proper exponentiation can be undertaken by working to two loop order , the
details of which will be published elsewhere) to get the functions $f(\Omega)$
and $a(\Omega)$ introduced in Eqs(3) and (5) . Note that since we are taking the logarithmic divergence for the bulk specific heat , the $C(\xi)$ in Eqns(4) and (5) reduces the constant C of Eqn(3) . The function $f(\Omega)$ and $a(\Omega)$ are
\bea
f(\Omega)=\frac{1}{2}(\frac{1}{-i\Omega}-1)\ln(1-i\Omega)
\eea
\bea
a(\Omega)=\frac{\pi}{2}\frac{1}{-i\Omega}[\sqrt{1-i\Omega}-1]
\eea
leading to
\ber
C(\xi,L,\omega)&=&C_0\{\ln \frac{\Lambda}{m}-\frac{1}{4}\ln(1+\Omega^2)-\frac
{1}{2\Omega}\tan^{-1}(\Omega)+i[\frac{1}{2}\tan^{-1}(\Omega) \nonumber \\
& & - \frac{1}{4\Omega}\ln (1+\Omega^2)]-\frac{\pi}{mL\Omega}(1+\Omega^2)^
{\frac{1}{4}}\sin(\frac{\tan^{-1}\Omega}{2})-\frac{i\pi}{mL\Omega} \nonumber \\
& &[(1+\Omega^2)^{\frac{1}{4}}\cos(\frac{\tan^{-1}\Omega}{2})-1]\} \nonumber \\
&=&C_R + i C_I
\eer
where $C_R$ and $C_I$ are the real and imaginary parts of the specific heat. 
Considering the zero frequency limit we see that $C_R = C_0[\ln \Lambda / m - \pi / 2mL]$
leading to $ a = \pi/2$ in Eq. (4) which is to be compared with $ a \simeq 1.4$ obtained
in [11].
\par
We now return to Eqn(1) , to find the attenuation and dispersion . The
{\bf{attenuation per wavelength}} is $\frac{\alpha\lambda}{2\pi}=\frac
{u_{1}C_{0}C_{I}}{u_{0}(C_R^{2} + C_I^{2})}$ which leads to the frequency 
attenuation $(\omega\gg 2\Gamma_{0} m^2)$ as
\bea
\frac{\alpha\lambda}{2\pi}=\frac{\pi u_1}{u_0}\frac{[1-2\sqrt{2}(\frac
{2\Gamma_0}{\omega L^2})^{\frac{1}{2}}]}{[\ln(\frac{\omega_0}{\omega})-\sqrt{2}
\pi(\frac{2\Gamma_0}{\omega L^2})^{\frac{1}{2}}]^{2}+\frac{\pi^{2}}{4}[1-
2\sqrt{2}(\frac{2\Gamma_0}{\omega L^2})^{\frac{1}{2}}]^{2}}
\eea
This is the {\bf{saturation attenuation}} per wavelength , which does not change as the temperature is lowered further , where $\omega_0/2\pi= 30 GHz, \Gamma_0 =1.2 \times 10^{-4} cm^{2}sec^{-1} , u_1/u_0 = 8/3\times 10^{-2}$ .  
\par
For the plate separation of $2110\AA$ of mehta and Gasparini , the reduction in the attenuation due to the quenching of fluctuations is about $18\%$ at $10 MHz$ and increases to $45\%$ at $2.5 MHz$ . This is a large effect compared to the $ 4\% $ surface effects that show up in the static measurements . For the corresponding measurement of thermal conductivity near the superfluid transition Kahn and Ahlers $[9]$ found that the deviation from the bulk is about $7\%$ when the correlation length $\xi$ equals the confining length $L$ (in their case the radius of the pore) . The surface effect for the ultrasonic measurement can easily amount to $30\%$ which makes this an attractive system for a confrontation between theory and experiment . The FSE on the dispersion can be obtained from the real part of Eqn(1).
\par
We note that the above is a one loop calculation in the critical region .
The lack of crossover to the background in our treatment of the specific heat implies that we can consider frequency $\omega$ which are much smaller than the cut-off frequency $\omega_0$ . This is a restriction on the validity of the dashed curve shown in Fig(1) . The solid curve in addition is restricted to confining lengths which are not too small i.e. $L\geq (2\Gamma_0/\omega)^{\frac{1}{2}}$ and in this regime the accuracy of the calculation is restricted by the loop order . This is not too severe a restriction as an accuracy of $O(\epsilon)$ which our calculation entails,becomes an accuracy of $O(\frac{\alpha}{\nu})$ when the combinatorial factors are included . Thus, in the above mentioned ranges of the parameters $\omega$ and $L$ , the dashed curve in Fig(1) should be an accurate prediction . It should be noted that contrary to the static specific heat or the thermal conductivity , the sound properties can only be probed in a real experiment . The final issue , then , is whether the effect can be observed in real a experiment . The critical ultrasonics near the superfluid transition has been studied more than two decades ago . The most accurate data of that period lies in the $0.5 MHz - 5 MHz$ range . In this region the scatter in the data is  about $15\%$. This is somewhat better than the borderline for detecting the suppression reported here . Considering the fact that developments in the experimental field would enable more accurate measurements at the present time , we believe this effect should be experimentally accessible .  
\par
The other sensitive part of an ultrasonic measurement is the low frequency end $(\omega\ll 2\Gamma_0 m^2)$ , where for the bulk substance the attenuation per wavelength is proportional to $C_R^{2}\Omega/4$ . The relative correction for the FSE is $1-\frac{\pi}{2mL}$ , once again a larger effect than can be obtained in statics. For an easily realizable situation of $ml\sim 8$ this gives a 20\% reduction in the attenuation . The whole course of the attenuation function with its dependence on $\omega$ and $L$ is straightforward to obtain and will be exhibited elsewhere . Here we have reported the salient feature , which carry the most experimentally accessible signatures . We hope this will stimulate experimental activity in the field.\\\\

$ {\bf {Acknowledgments}} $  \\\\\
One of the authors (SB) would like to thank the C.S.I.R, India, for providing 
partial financial support and Dr.Manabesh Bhattacharya for his help and encouragement.
We thank the referee for pointing out various improvements.

\newpage
$ \bf {Figure\:\:caption} $ \\\\

Fig.1. Saturation attenuation is plotted against frequency.The dashed curve
shows the bulk ($L\rightarrow\infty$) result whereas the solid curve shows the surface effect. \\


\newpage
\begin{thebibliography}{99}
\bibitem{1} For a review see, V.Dohm, Physica Scripta
\bibitem{2} S. Mehta and F. M. Gasparini, Proceedings of LT-21  Czech. J.Phys 
{\bf{46}} Suppl S1  173 (1996) ,Phys. Rev. Lett {\bf{78}} 2596  (1997)
\bibitem{3} J. P. Chen and F. M. Gasparini, Phys. Rev. Lett {\bf{40}} 331 (1970) 
\bibitem{4} F. M. Gasparini , G. Agnolet and J. D. Reppy, Phys. Rev. {\bf{B29}}
 138  (1984)
\bibitem{5} I. Rhee , F. M. Gasparini and D. J. Bishop, Phys. Rev. Lett.
{\bf{63}}  410 (1989) 
\bibitem{6} J. A. Nissev, T. C. P. Chui and J. A. Lipa, J.Low Temp. Phys
{\bf{92}} 393 (1993)
\bibitem{7} M. Coleman and J. A. Lipa, Proceedings of LT-21, Czech. J. Phys.
{\bf{46}}  S1  183 (1996)
\bibitem{8} G. Ahlers and R. V. Duncan, Phys. Rev. Lett.{\bf{61}}  846  (1988)
\bibitem{9} A. M. Kahn  and  G. Ahlers, Phys. Rev. Lett {\bf{73}}  944  (1995) 
\bibitem{10}W. Huhn and  V. Dohm , Phys. Rev. Lett {\bf{61}}  1368  (1988)     
\bibitem{11}R. Schmolke, A. Wacker, V. Dohm and D. Frank , Physica {\bf{B165}} 
 {\bf{166}} 575  (1990)
\bibitem{12}P. Sutter and V. Dohm , Physica{\bf{B  }}        (1993)
\bibitem{13}A. B. Pippard, Phil. Mag. {\bf{1}}  473  (1956)
\bibitem{14}M. J. Buckingham  and  W. M. Fairbank , Prog. in Low Temp. Phys.
,ed. by C. J. Gorter (North-Holland , Amsterdam ) {\bf III}  80  (1961) 
\bibitem{15}R. A. Ferrell and J. K. Bhattacharjee , Phys. Rev. Lett {\bf{44}}
403  (1982)  
\bibitem{16}R. A. Ferrell and J. K. Bhattacharjee , Phys. Rev. {\bf{B23}}  2484
(1981)   
\bibitem{17}J. Pankert and V. Dohm , Phys. Rev.{\bf{B40}} 10856  (1989)


\end{thebibliography}
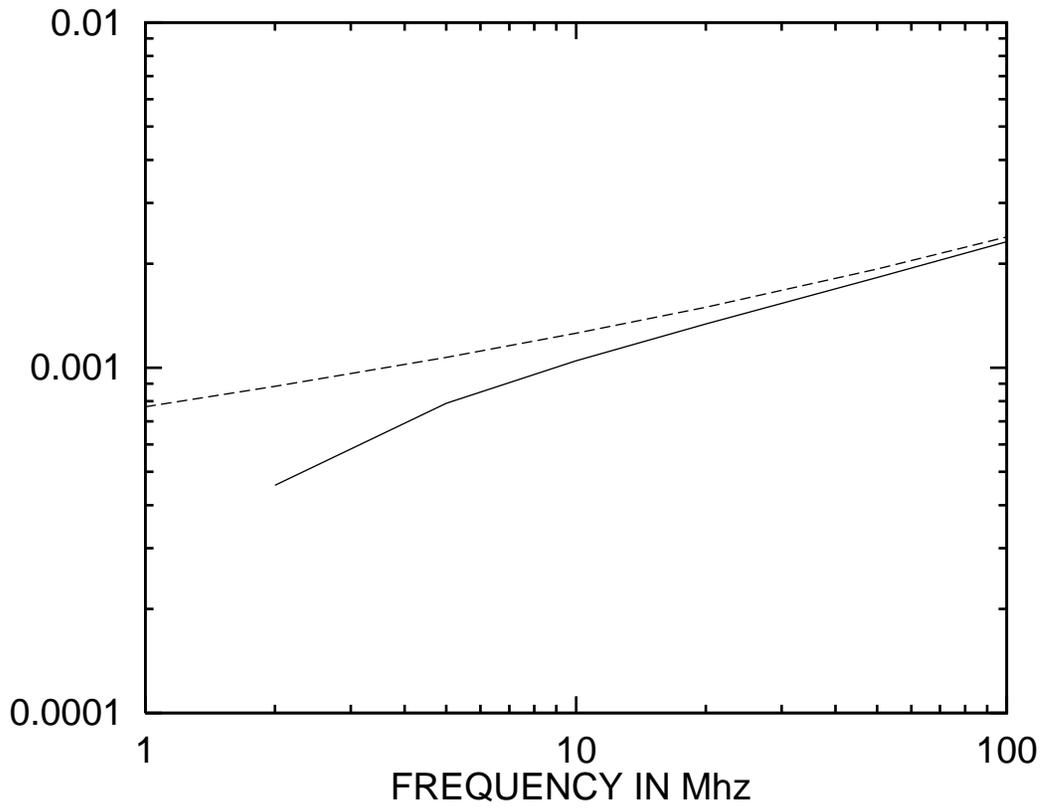
\end{document}